\documentclass[reprint,amsmath,amssymb,aps,prl,showpacs,floatfix]{revtex4-1}

\usepackage{graphicx}
\usepackage{dcolumn}
\usepackage{bm}
\usepackage{hyperref}
\usepackage{subfigure}
\usepackage{extraplaceins}
\begin{document}


\title{Y-junction splitting spin states of moving quantum dot}

\author{Tuukka Hiltunen}
\author{Ari Harju}

\affiliation{%
COMP Centre of Excellence, Department of Applied Physics, Aalto University, Helsinki, Finland
}%

\date{\today}

\begin{abstract} 
The development of a working quantum computer utilizing electron spin states as qubits is a major goal for many theorists and experimentalists.
The future applications of quantum information technology would also require a reliable method for the transportation of quantum information. 
A promising such method is the propagation of electrons by a surface acoustic wave (SAW). In this paper, we simulate the SAW transfer of two
interacting electrons through a Y-shaped junction. Our results show that the singlet and triplet states
can be differentiated in the Y-junction by an electrostatic detuning potential, an effect that could be used in for example measuring the state
of a two-spin qubit. 
\end{abstract}

\maketitle

\noindent The spins of electrons are promising candidate for a qubit \cite{loss}. A framework for using two-electron spin eigenstates as qubits was proposed by Levy in 2002 \cite{levy}.
After that, a lot of progress has been made on this direction \cite{spins}. 
The coherent electrical control and measurement of two-spin qubits has been presented experimentally \cite{petta}. Microscopic theory \cite{sarkka} can explain experiments \cite{petta2}
and predict new phenomena \cite{studenkin}.
Remarkably long dephasing times in two electron spin states have been shown \cite{bluhm}. Recently, entanglement between a pair of two-spin qubits has been
demonstrated experimentally \cite{shulman}. This is an important achievement, as entanglement is essential in a working quantum computer. 

The two-electron spin eigenstates are the singlet state and the triplet states.
The different symmetry properties of the spatial wavefunctions affect the behavior of the electrons. Electrons in the singlet behave effectively as bosons. The so called exchange force, a geometrical consequence of the symmetrization of the spatial wave function,
causes an attraction between the electrons in the singlet state whenever their wave functions overlap. For the triplet state the effect is the opposite, the overlap of the wave functions results in
an effective repulsive force between the electrons. This phenomenon is important for the quantum information technology and we show that it can be used to differentiate the singlet and triplet states or for example to measure the spin state of the system.

Few-electron circuits of quantum information technology, consisting of networks of quantum dots, require a method for transportation of electrons between different parts of the circuit.
A very promising method for transferring electrons between quantum dots is by a surface acoustic wave (SAW). A framework for SAW transport based quantum computing was presented by Barnes et al. \cite{barnes}. 
It has been shown experimentally that single electrons can be captured from one quantum dot and transported into another by a SAW pulse moving along a one dimensional channel \cite{saw3,saw4}.
The SAW induced transfer of electrons has also been realized in the cases of Y- and X-shaped junctions \cite{sawy,kataoka}.

\begin{figure}[h]
\vspace{0.3cm}
\includegraphics[width=.70\columnwidth]{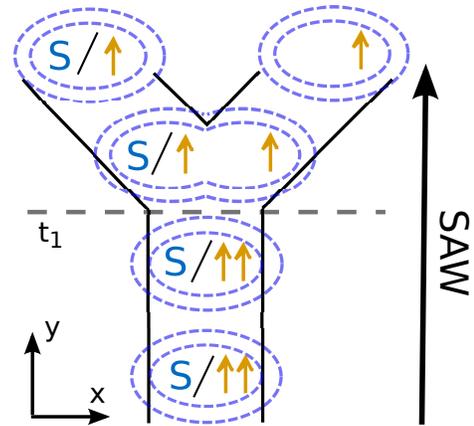}
\caption{(Color online) SAW induced transfer of electrons through an Y-shaped junction. The potential minima caused by the SAW pulse are modeled as elliptic quantum dots moving
at a constant velocity. The initial dot is split into two similar dots when it meets the Y-junction at the time $t=t_1$. 
The singlet state (denoted by S) is transferred to the Y-branch that is energetically preferable due to a detuning potential, while the triplet state (denoted by two
spin-up arrows) is split into the two Y-branches.} 
\label{fig:yjunction}
\end{figure}

In this letter, we simulate a system of two electrons moving along a SAW pulse in a Y-shaped channel in which the Y-branches are detuned using electrostatic gating.
The potential minimum created by a SAW pulse can be thought as a moving quantum dot that confines the electrons during their transfer \cite{barnes,sawy}.
As the SAW pulse moves at a constant velocity, the system can be modeled as a stationary quantum dot that splits into two separate dots at the Y-junction, see Fig. \ref{fig:yjunction}.
We show that, due to the exchange force, the Y-junction and the detuning potential can be used to split the electrons in the triplet states into the two different Y-branches,
while the electrons in the singlet state are both transferred into the Y-branch that is energetically more preferable due to the detuning.

We model the SAW transfer of a two-electron system through a Y-junction with the time-dependent Hamiltonian
\begin{equation}\label{eq:ham}
H(t)=\sum_{j=1}^2\left[-\frac{\hbar^2}{2m^*}\nabla_j^2+V_{ext}(\textbf{r}_j,t)\right]+V_{int}(\textbf{r}_1,\textbf{r}_2).
\end{equation}
Here, $V_{int}$ is the pairwise Coulomb interaction, and $m^*$ is the effective mass of an electron. The magnetic field is omitted in our computations.

The external confinement is initially an elliptical quadratic well centered at the origin.
At time $t=t_1$, the quantum dot is split into two dots that start to move apart from each other. The external confinement potential $V_{ext}(\textbf{r}_j,t)$ is written as
\begin{equation} \label{eq:vext}
\frac{1}{2}m^*\omega_0^2\left(\alpha\min\{|x_j-a(t)/2|^2,|x_j+a(t)/2|^2\}+\beta y_j^2\right)+Dx_j.
\end{equation}
Here, $a(t)$ is the distance of the dot minima. $a(t)=0$ when $t<t_1$ and when $t\geq t_1$ it is an increasing function of $t$. A linear electrostatic detuning potential of strength $D$ is included. With $D=0$, the minima of the dots are located at points $(x,y)=(\pm a(t)/2,0)$. A non-zero $D$ shifts the locations of the minima slightly (their distance is still $a(t)$). In Eq. (\ref{eq:vext}), $\alpha$ and $\beta$ define the $x$- and $y$-direction confinements of the dot.
A schematic of the Y-junction can be seen in Fig. \ref{fig:yjunction}

The ground state of (\ref{eq:ham}) is computed by using the exact diagonalization (ED) method. The Fock-Darwin (FD) states are used as
the one-particle basis in the ED.
The matrix elements for the Coulomb interaction and the parabolic external confinement
can be computed analytically for the FD-states. The Hamiltonian is
stored as a sparse matrix and its ground state energy and eigenvector are computed by the Lanczos-iteration. The time evolution of the system is computed by propagating the initial ground state $\psi(0)$.
\begin{equation}
\psi(t+\Delta t)=\exp\left(-\frac{i}{\hbar}H(t)\Delta t\right)\psi(t),
\end{equation}
where the matrix exponential is evaluated using the Lanczos algorithm. The charge density was computed from the two-body wave function using the reduced one-particle
density matrix.

The two-body wave function, a product of the spin- and spatial parts, is anti-symmetric. A symmetric spin part corresponds to an anti-symmetric spatial wave function, and vice versa. 
The four lowest eigenstates of a two-electron system are the spatially symmetric singlet-state $|S\rangle$, and the spatially anti-symmetric triplet-states $|T_-\rangle$, $|T_0\rangle$, and $|T_+\rangle$. In the zero magnetic field,
the three triplet states are degenerate
and the singlet is always the ground state. In the Lanczos-method, the triplet-states can be obtained by fixing the $z$-component of spin to $\pm1$.

In our simulations, the confinement strength was set to $\hbar\omega_0=3$ meV. Several different shapes of elliptical dots were used (the values of $\alpha$ and $\beta$ in Eq. (\ref{eq:vext})). 
The typical GaAs parameter values for the effective mass and the dielectric constant are $m^*=0.067m_e$ and $\epsilon_r=12.7$.
After the split at $t=t_1$, $a(t)$ was increased linearly from zero to the value of $100$ nm. 
The linear detuning strength $D$ was set to be in the same order as the linear part of the parabolic confinement, $D\sim D_0=m^*\omega_0^2\times30$ nm. In experiments,
this would correspond to two gates with a potential difference of approximately 50 mV set 200 nm apart from each other.
The many-body Hamiltonian was created using the first 120 FD-states in energy. The convergence
of the singlet- and triplet-energies was studied with respect to the basis size, and 120 FD-states was found to be enough to obtain accurate results up to the distance 100 nm between the dot minima. 

We did the simulations using several different time windows and detuning strengths. The time step $\Delta t$ was set to $\Delta t \leq 1$ ps.
This was found to be small enough to ensure accurate computation of the dynamics. It should be noted that the simulation
can be started at $t=t_1$, as the system is in the ground state at $t<t_1$. The relevant timescale in the SAW transfer through our simulated system is from 10 ps to the nanosecond scale \cite{saw3,saw4,kataoka}.

\begin{figure}[!t]
\vspace{0.3cm}
\includegraphics[width=\columnwidth]{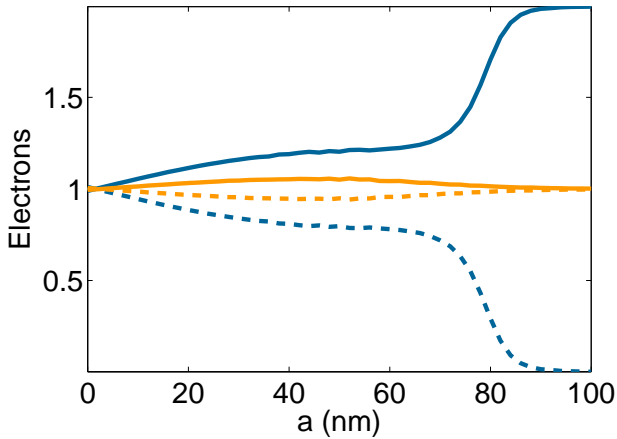}
\caption{(Color online) Number of electrons in the left- (solid lines) and right-dot (dashed lines) as a function of the dot distance $a$. Singlet (blue/dark grey) and triplet (orange/light grey) results are both shown in the figure. The detuning strength was set to $D=0.20\times m^*\omega_0^2\times30$nm, and the time
window is 1 ns.
}
\label{fig:electrons}
\end{figure}

\begin{figure}[!tb]
\vspace{0.3cm}
\subfigure{\includegraphics[width=.48\columnwidth]{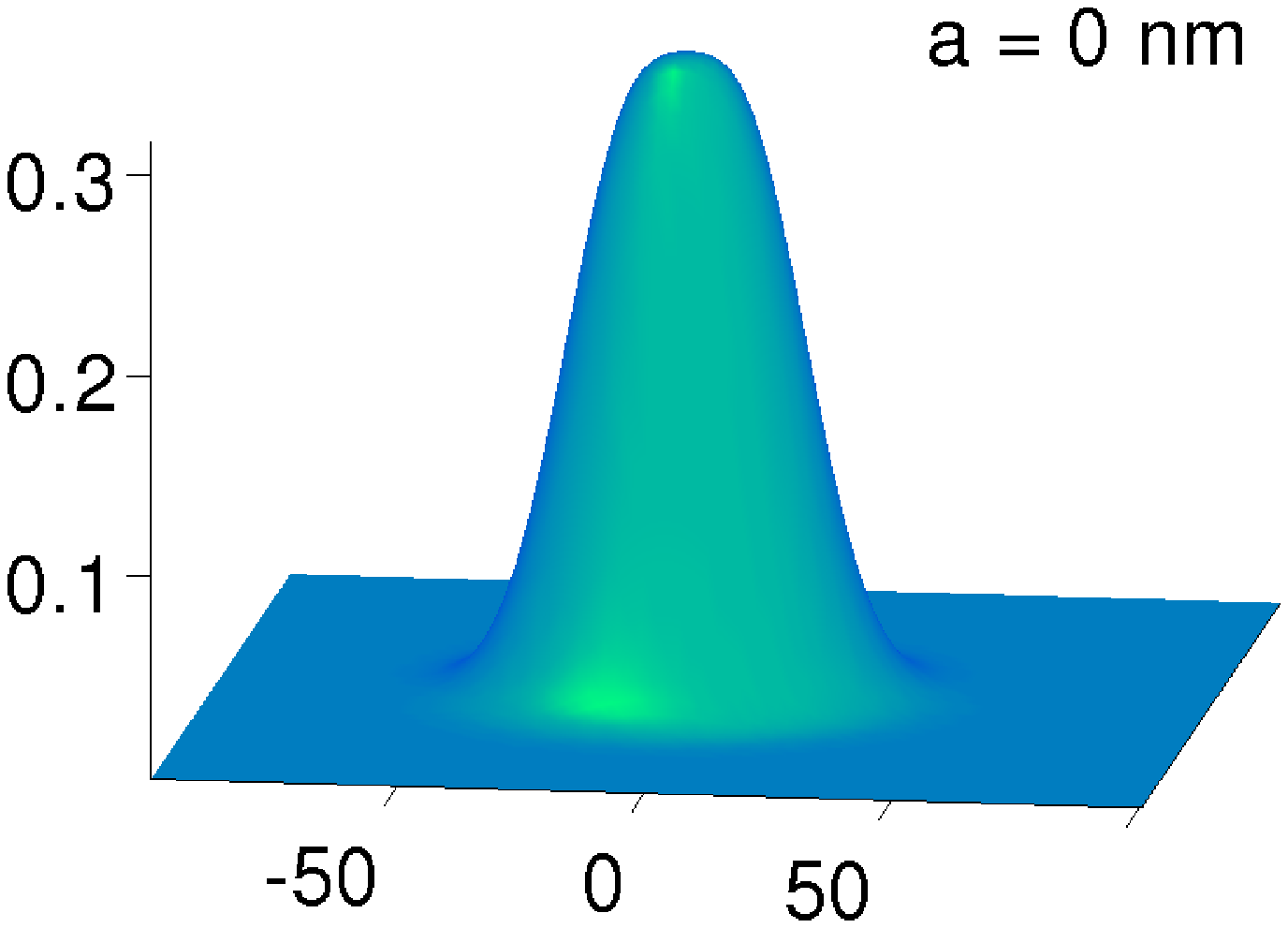}}
\subfigure{\includegraphics[width=.48\columnwidth]{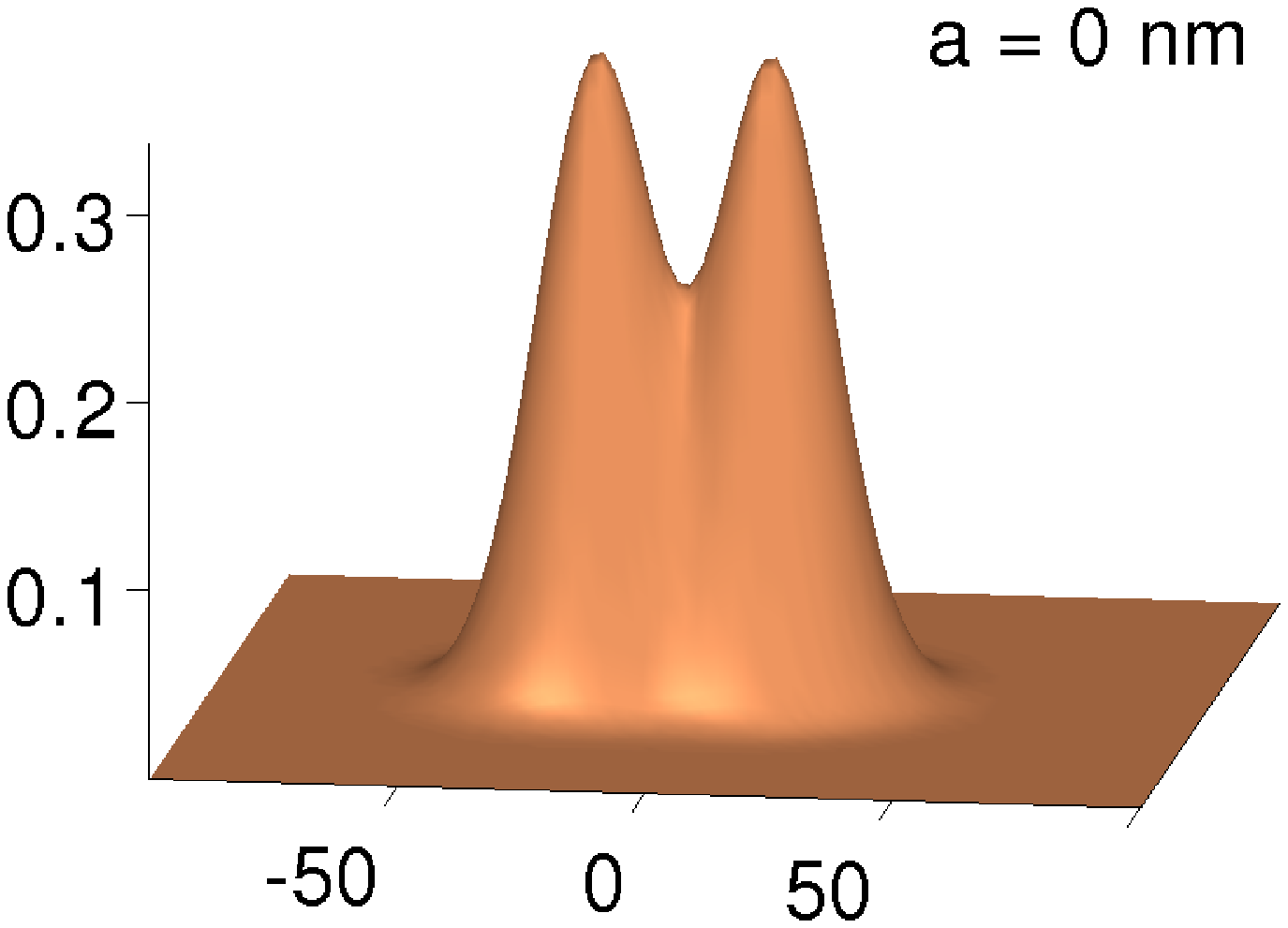}}
\subfigure{\includegraphics[width=.48\columnwidth]{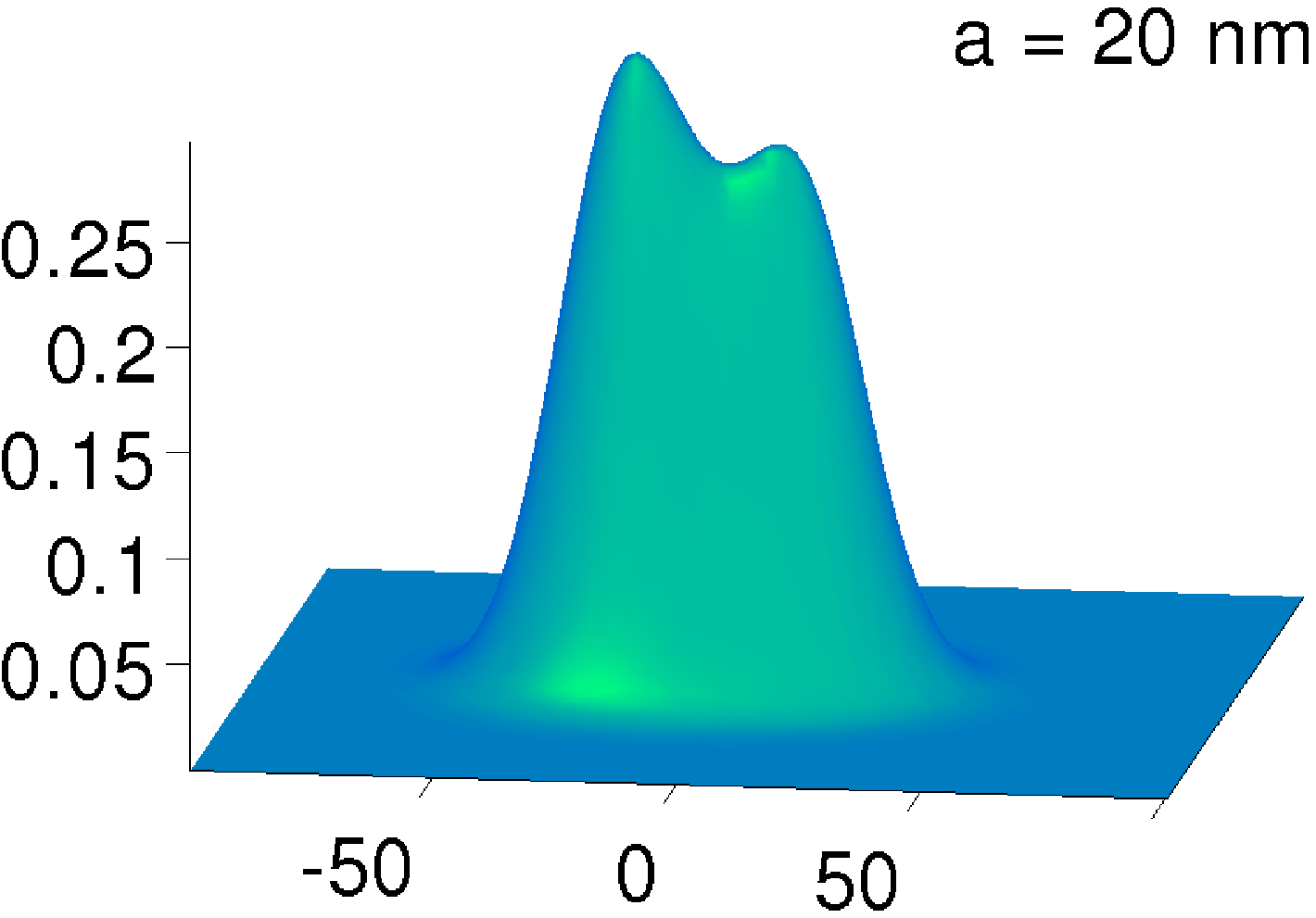}}
\subfigure{\includegraphics[width=.48\columnwidth]{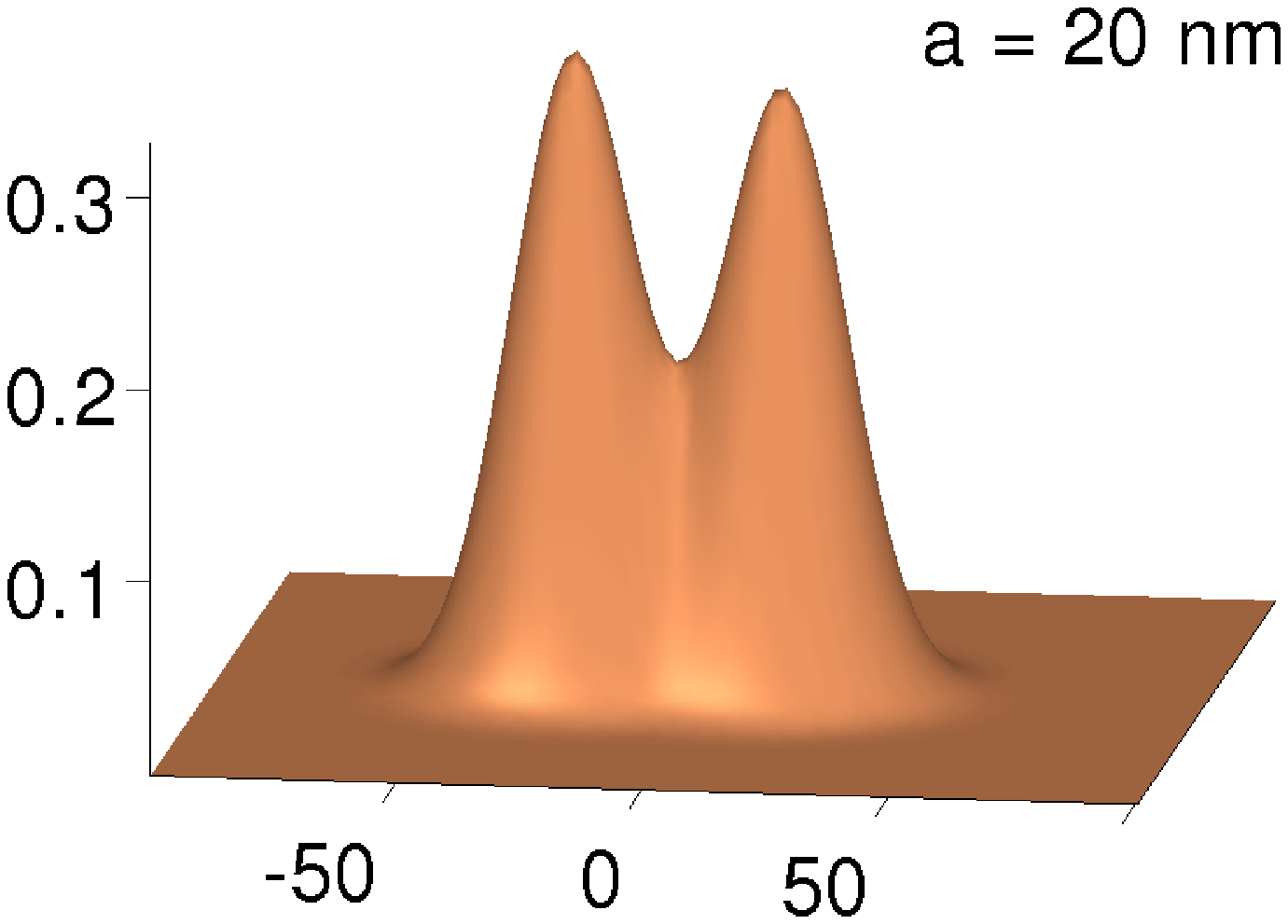}}
\subfigure{\includegraphics[width=.48\columnwidth]{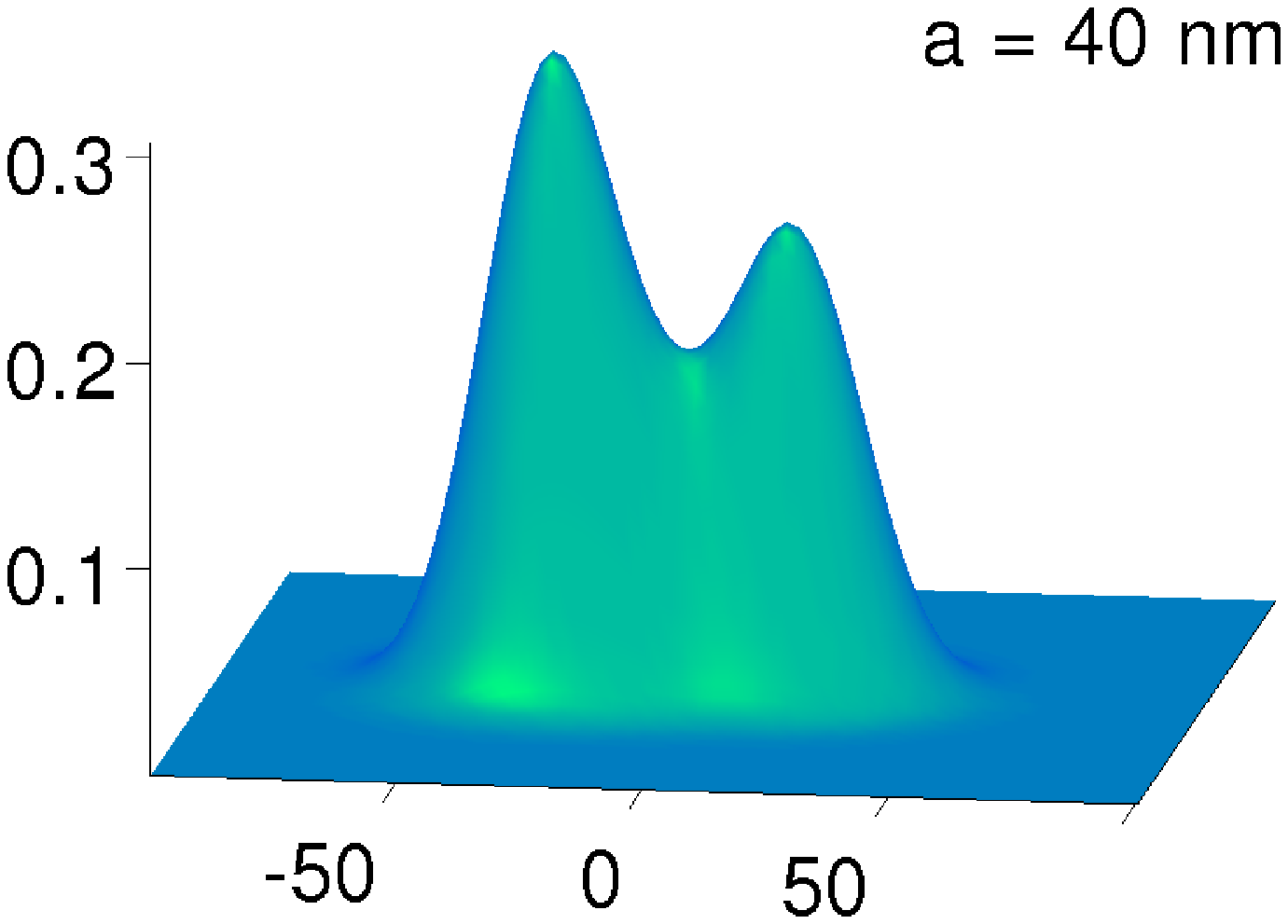}}
\subfigure{\includegraphics[width=.48\columnwidth]{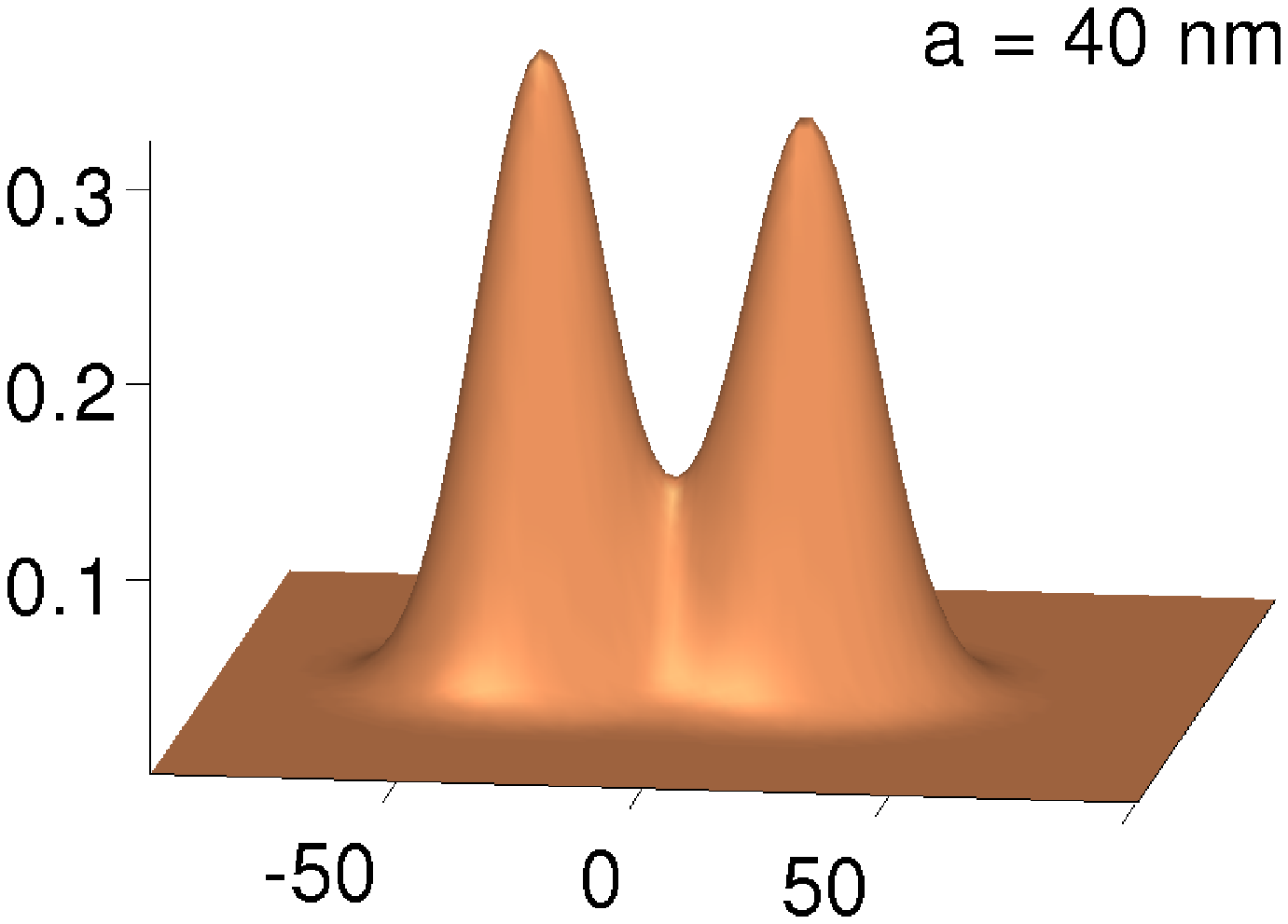}}
\subfigure{\includegraphics[width=.48\columnwidth]{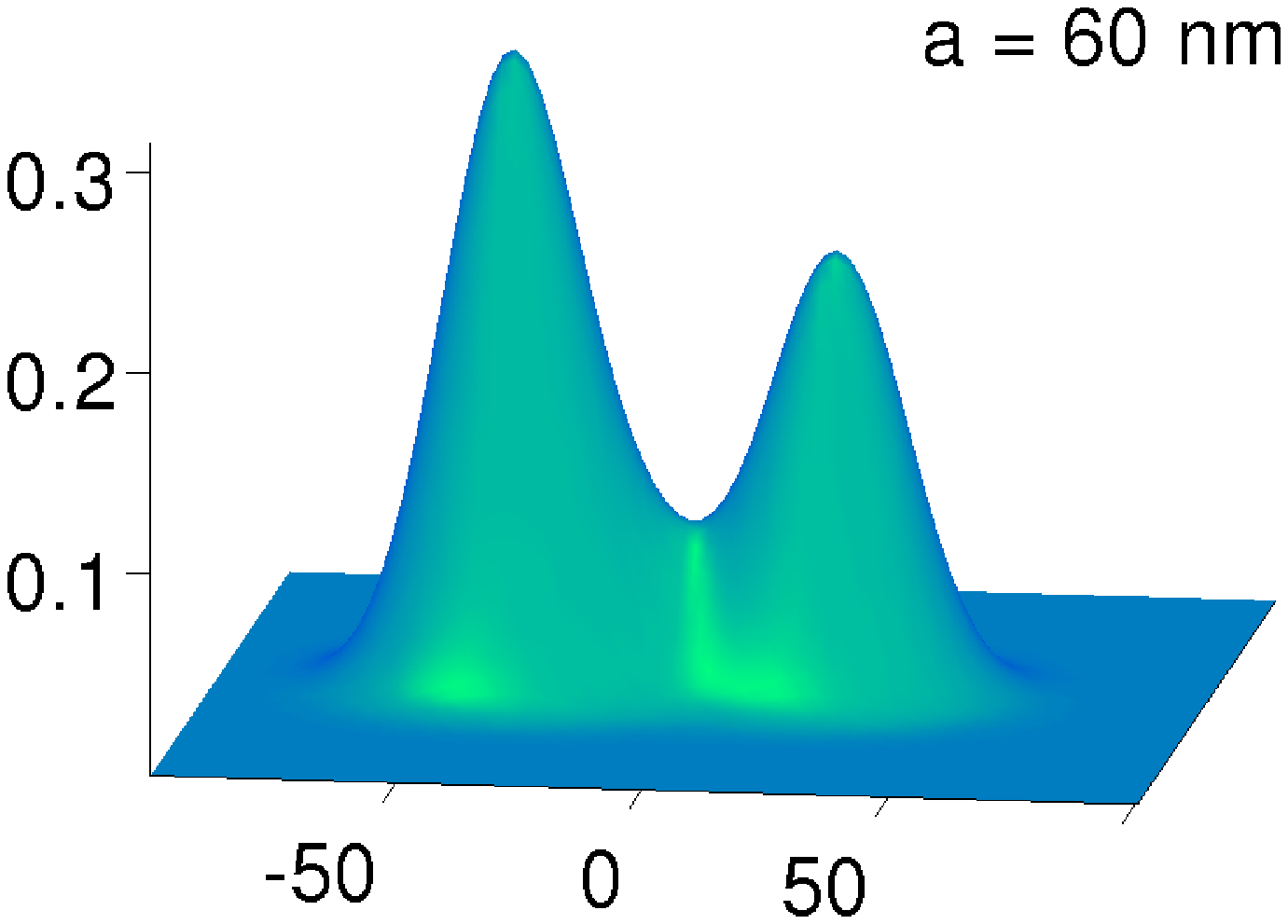}}
\subfigure{\includegraphics[width=.48\columnwidth]{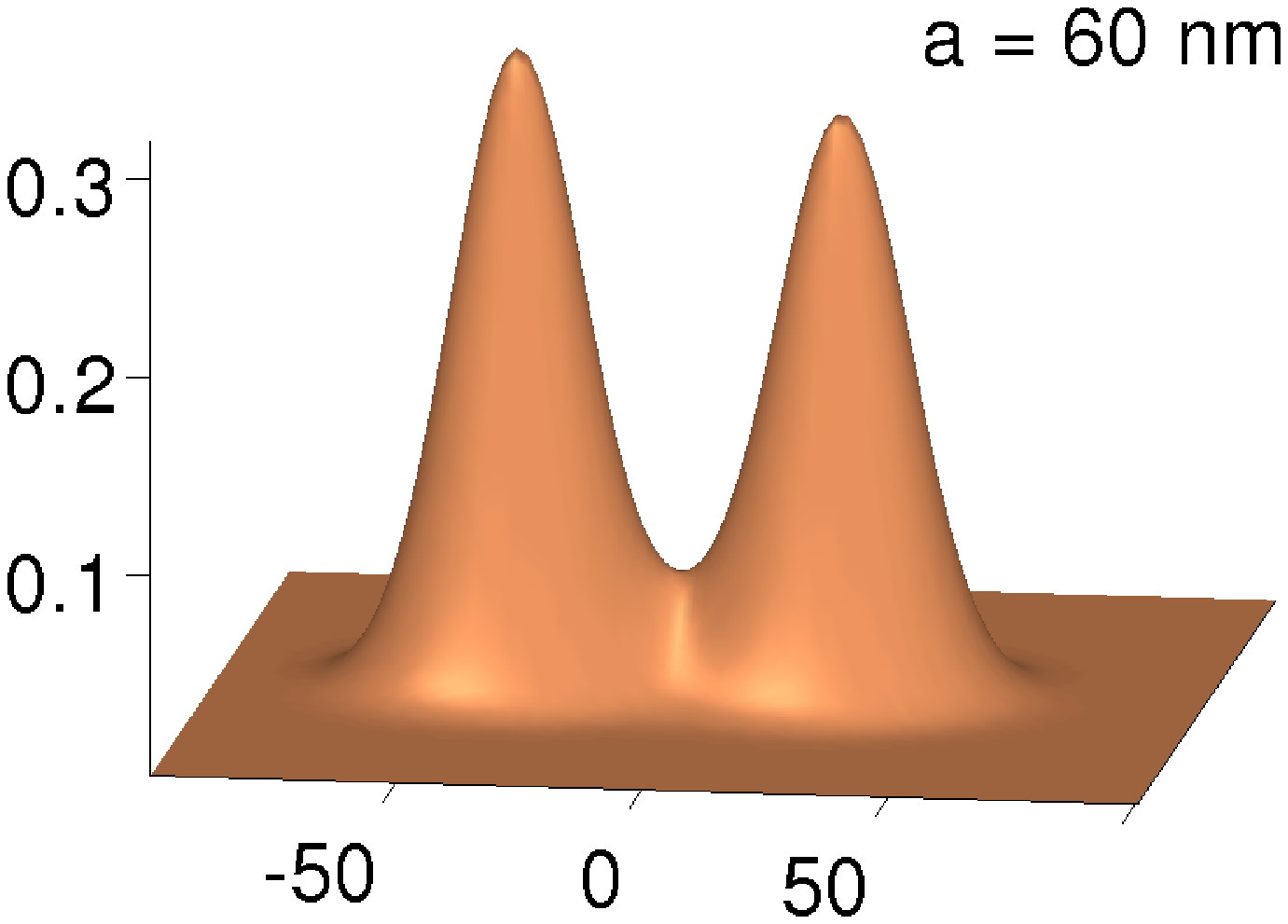}}
\subfigure{\includegraphics[width=.48\columnwidth]{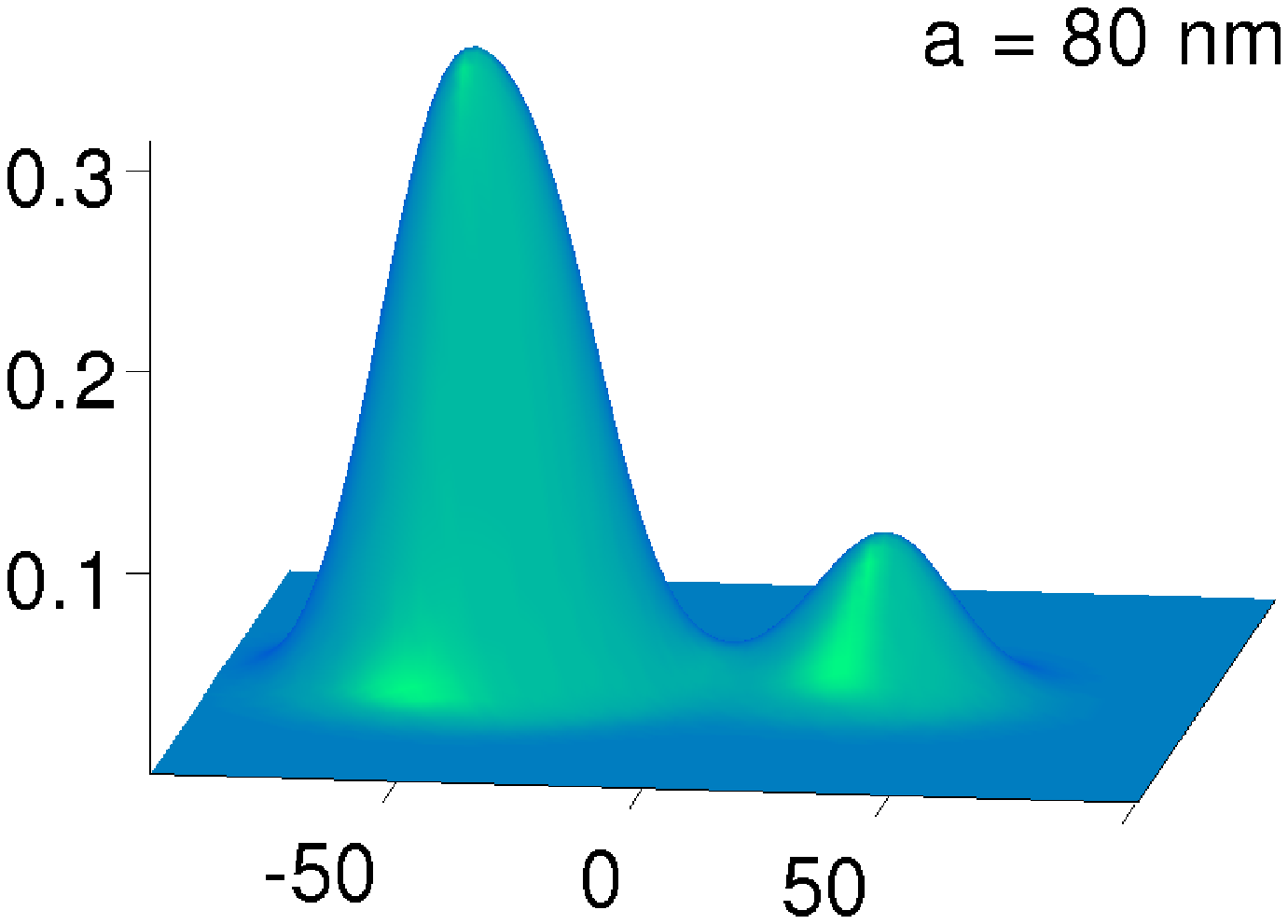}}
\subfigure{\includegraphics[width=.48\columnwidth]{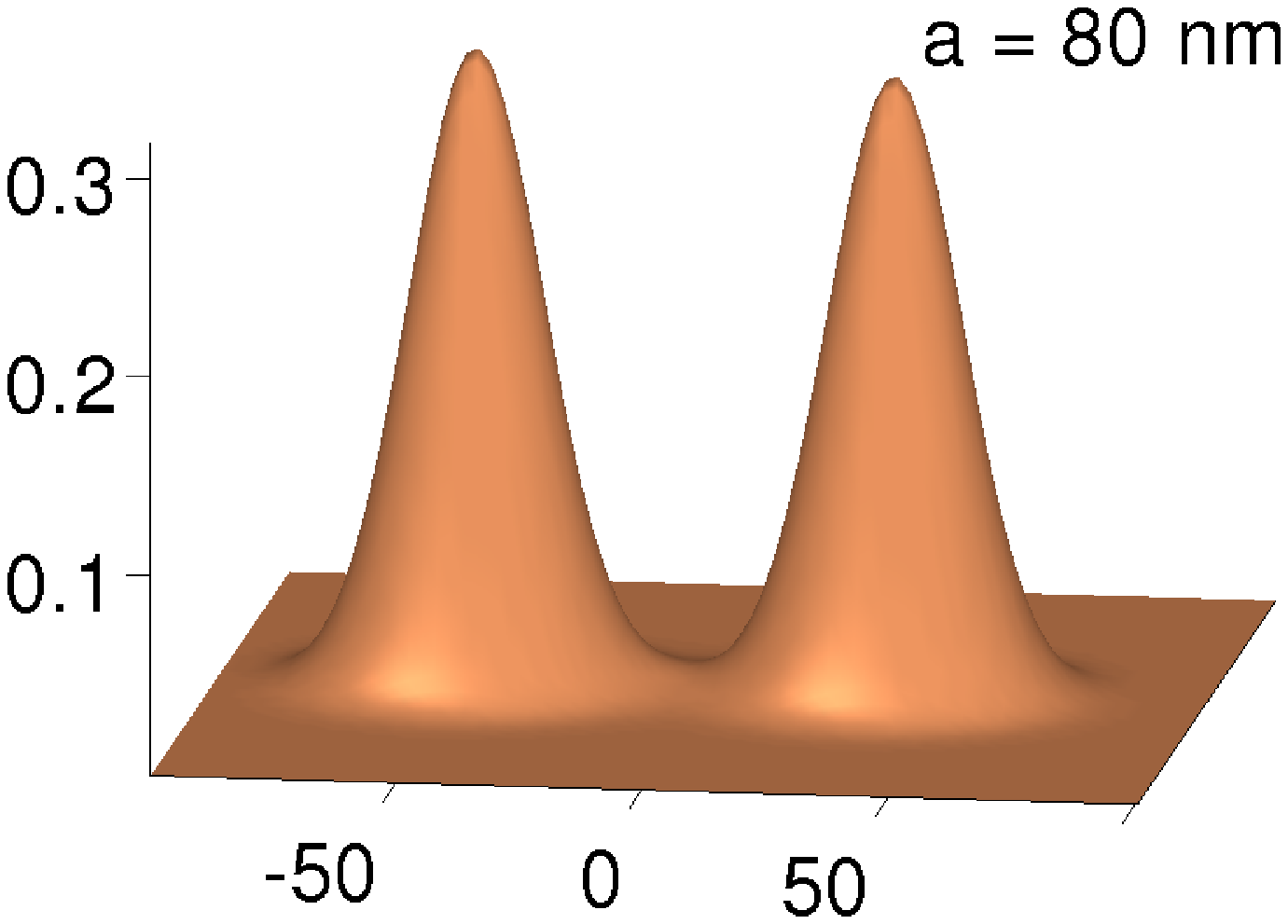}}
\subfigure{\includegraphics[width=.48\columnwidth]{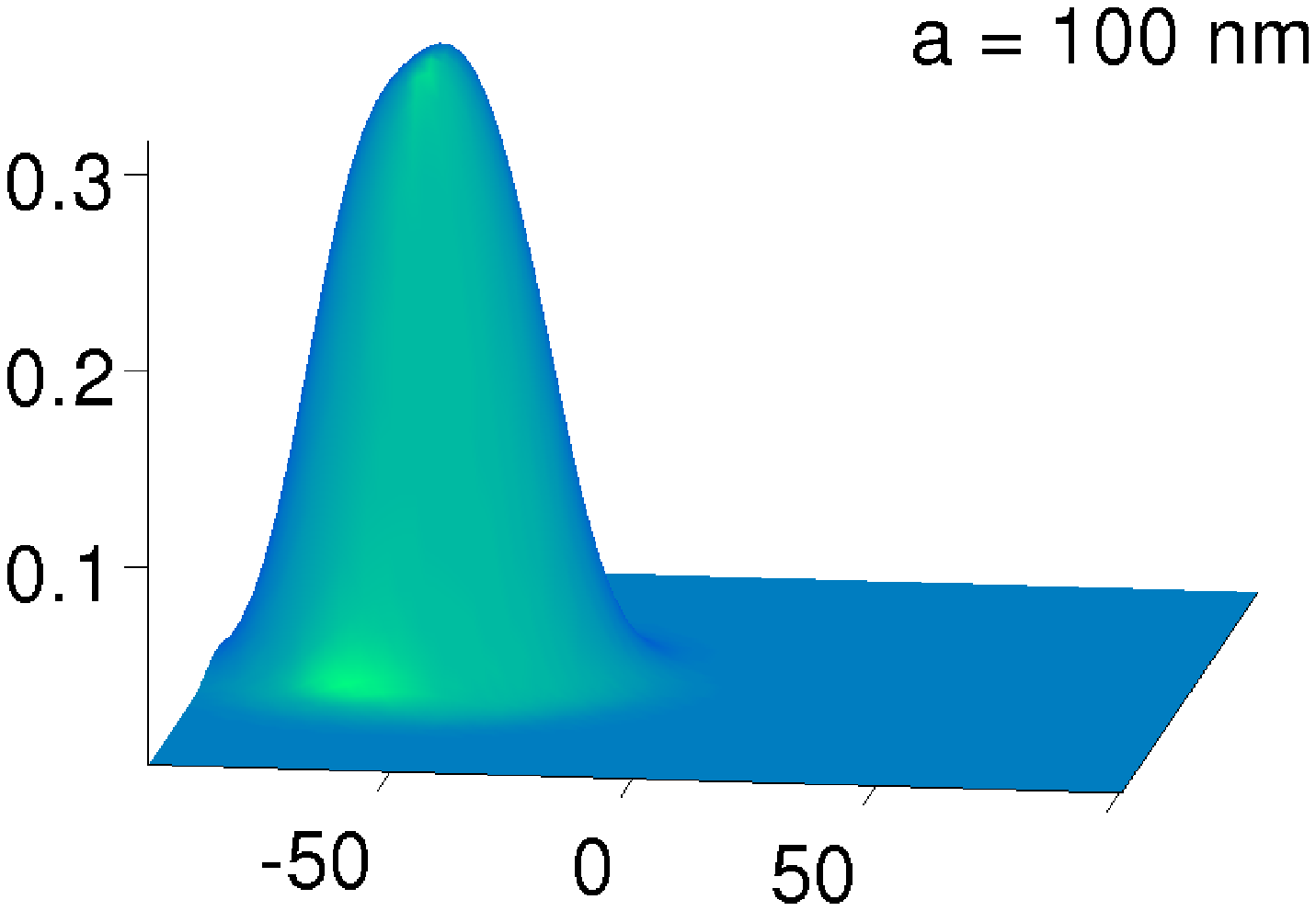}}
\subfigure{\includegraphics[width=.48\columnwidth]{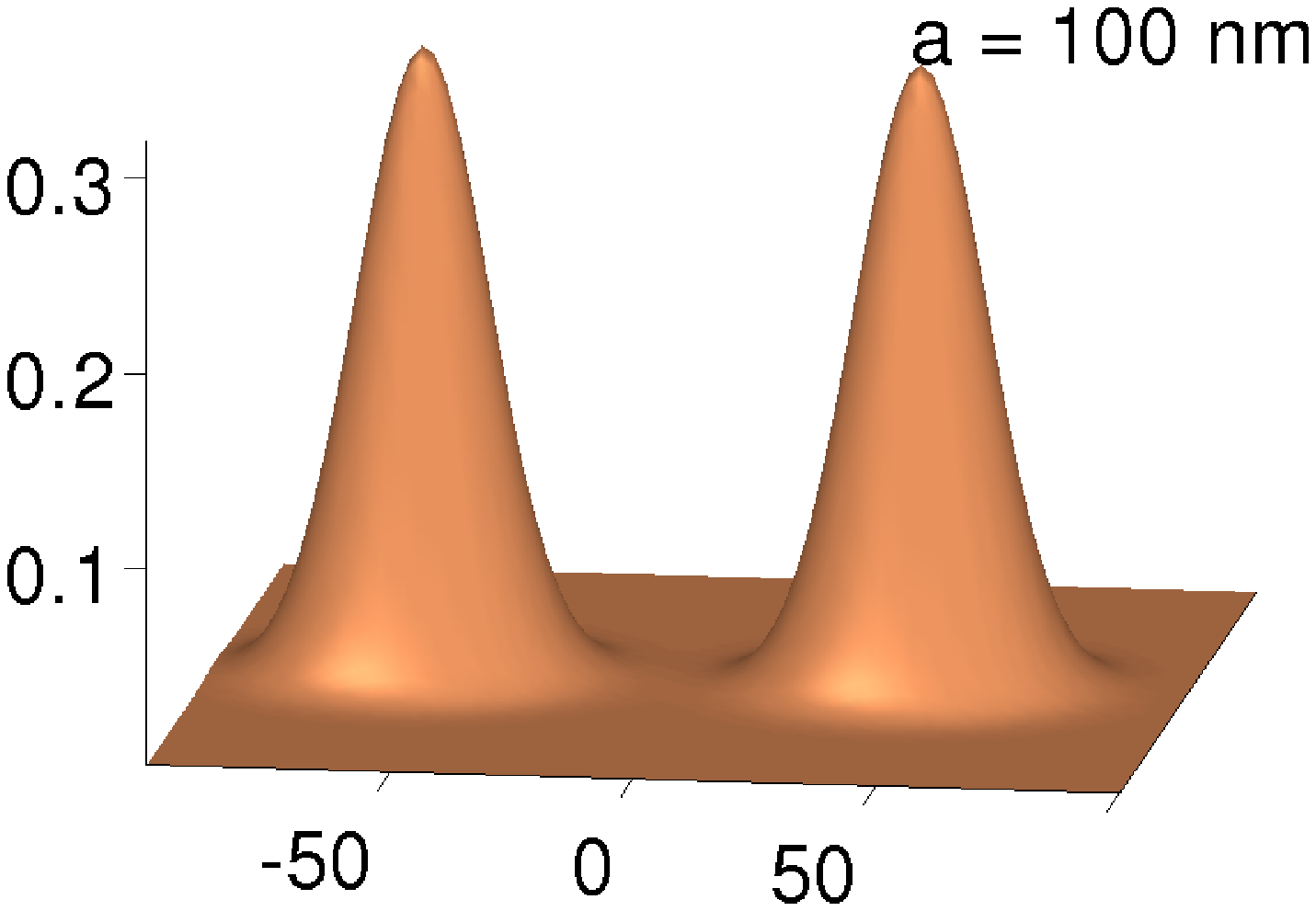}}
\caption{(Color online) The evolution of the charge density as $a$ is increased from 0 nm to 100 nm. The density is shown at 20 nm intervals, starting from $a=0$. The blue plots in the left show the density of the singlet state, and the
orange plots in the right the triplet state(s). The detuning strength was set to $D=0.20\times m^*\omega_0^2\times30$nm, and the time
window is 1 ns.}
\label{fig:density}
\end{figure}

The results of our simulation in which $a$ is increased form 0 to 100 nm in the time of 1 ns can be seen in Figs. \ref{fig:electrons} and \ref{fig:density}. The figures were obtained with approximately spherical dots, $\alpha=0.9999$ and $\beta=1$. We included a small radial asymmetry
to align the electrons along the $x$-axis in the initial dot. The detune strength is set to be $D=0.20\times m^*\omega_0^2\times30$nm. The number
of charges in the left and right dots are shown as a function of $a$ in Fig. \ref{fig:electrons}. The number of electrons in the left dot
was obtained by integrating numerically the charge density in the half plane $x<x_0$, where $x_0={-D}/{m^*\omega_0^2}$ is the middle point between the dots.  As for the right dot, the integration was done in the area $x\geq x_0$.
Fig. \ref{fig:density} shows the evolution of the electron density during the simulation. The densities for both the singlet- and triplet-states are shown at 20 nm intervals, starting
from $a=0$ nm.

The singlet electron density in Fig. \ref{fig:density} shows initially just one large central peak. 
The triplet density shows two very distinct peaks corresponding to the two electrons even at $a=0$ when there is just one potential minimum. This difference is explained by the symmetry properties of the many-body wave functions; the anti-symmetry of the spatial wave function of the triplet
state results in the density vanishing when the electrons are near to each other, $\textbf{r}_1\approx\textbf{r}_2$.

As the dots are pulled apart from each other, the singlet-electrons start to localize more into the left dot, as can be seen in Fig. \ref{fig:electrons}.
The density in Fig. \ref{fig:density} shows that at first a second peak starts to form in the right dot. However, the amplitude of this second peak decreases (compared to
the left peak) as the distance of the dots increases. The right peak disappears when the distance of the dots is about 80 nm. When $a=100$ nm, both electrons are fully localized in the
left dot, as can be seen in Figs. \ref{fig:electrons} and \ref{fig:density}.

The triplet state behaves in a qualitatively different way as the dots are pulled apart. The density peak in the right dot
never vanishes, it just moves further right. At $a=100$ nm, the density goes to zero between the dots, and there is one electron in the left dot and one in the right. The electron density is symmetric
with respect to the two dots, as can be seen in the integrated electron density in Fig. \ref{fig:electrons}.
The spatial antisymmetry and the resulting repulsive exchange force does not allow both the triplet electrons inhabiting the energetically preferable left dot.

The rate at which $a$ is increased was found to affect the results. If $a$ grows slowly, the time evolution is approximately adiabatic; at the time $t$ the system is
in the ground state of Eq. (\ref{eq:ham}). The results shown in Figs. \ref{fig:electrons} and \ref{fig:density}  (the time window is 1 ns) are very close to the adiabatic behavior of the system.
When the time window is decreased to the value of 10 ps, the singlet state is no longer transferred as completely into the left branch as in Fig. \ref{fig:density}. A small but visible peak stays
in the right dot even at $a=100$ nm. As the window is further decreased, more and more charge is left in the right dot at the end of the simulation. It should be noted that in our simulations,
the time window length corresponds to both the SAW pulse speed and the angle of the Y-junction. The higher the angle (relative to the $y$-axis in Fig. \ref{fig:yjunction}) is
the faster the distance between the dots at the two Y-branches increases.

\begin{figure}[!t]
\vspace{0.3cm}
\subfigure{\includegraphics[width=.48\columnwidth]{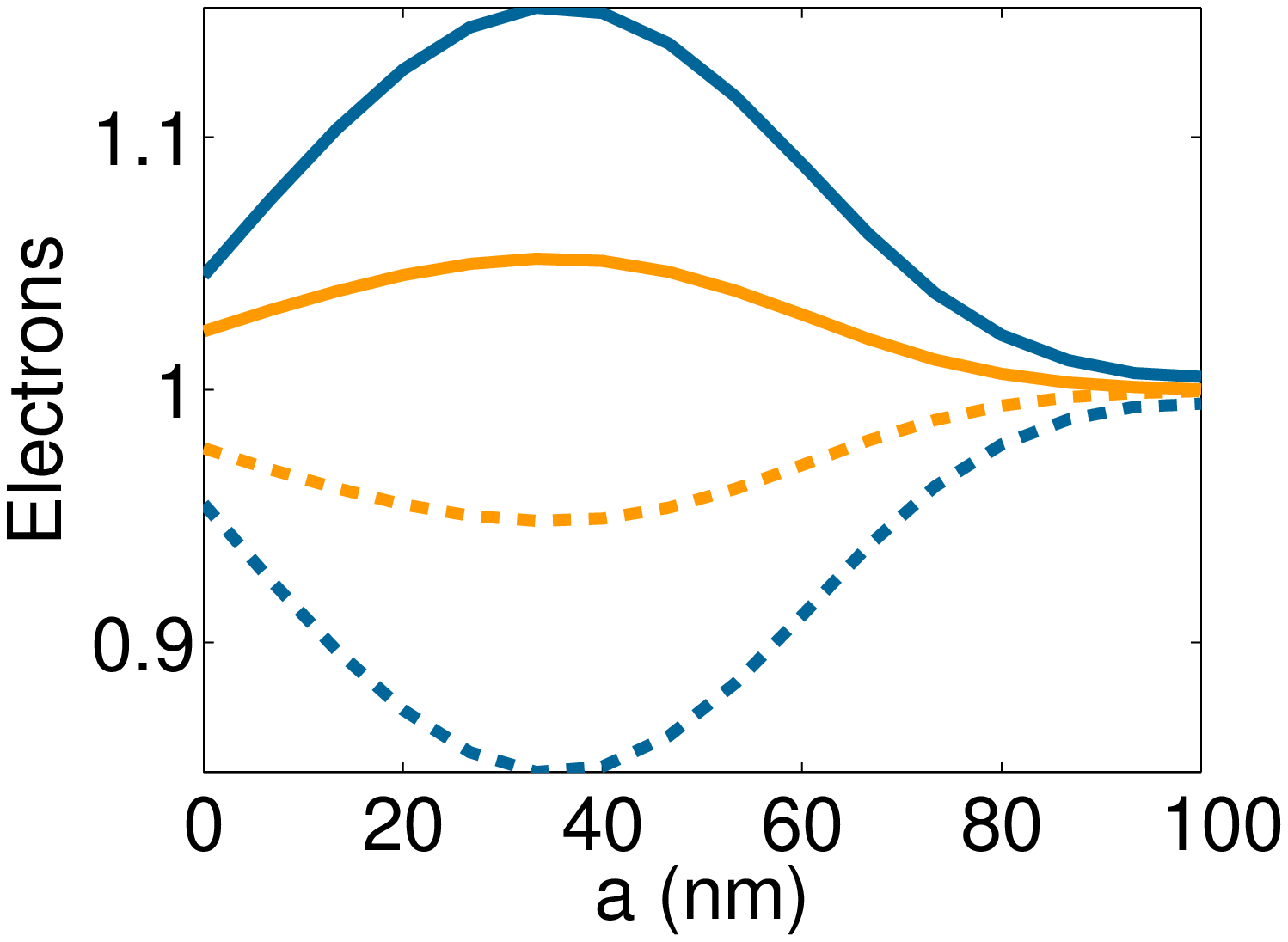}}
\subfigure{\includegraphics[width=.48\columnwidth]{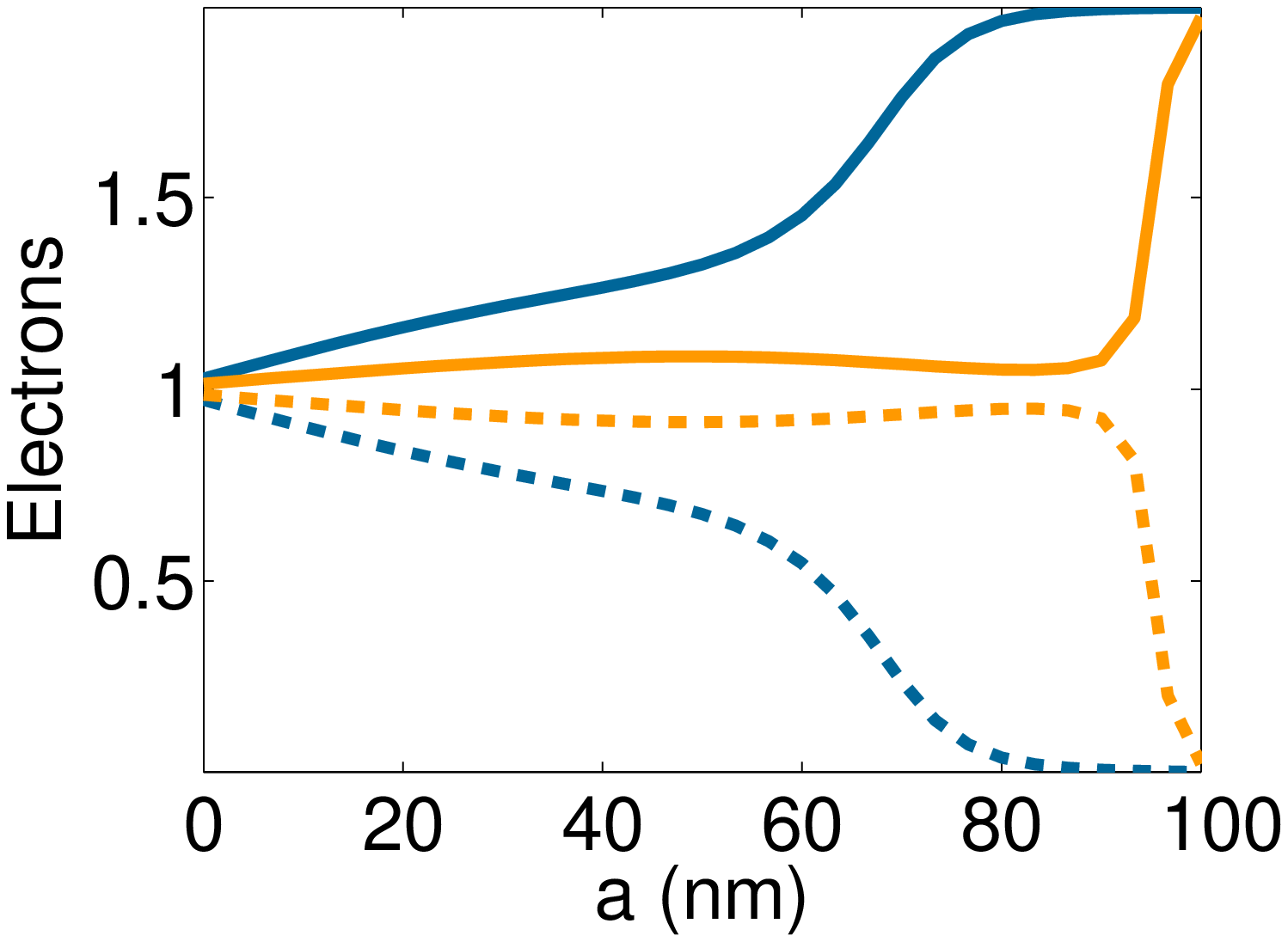}}
\subfigure{\includegraphics[width=.48\columnwidth]{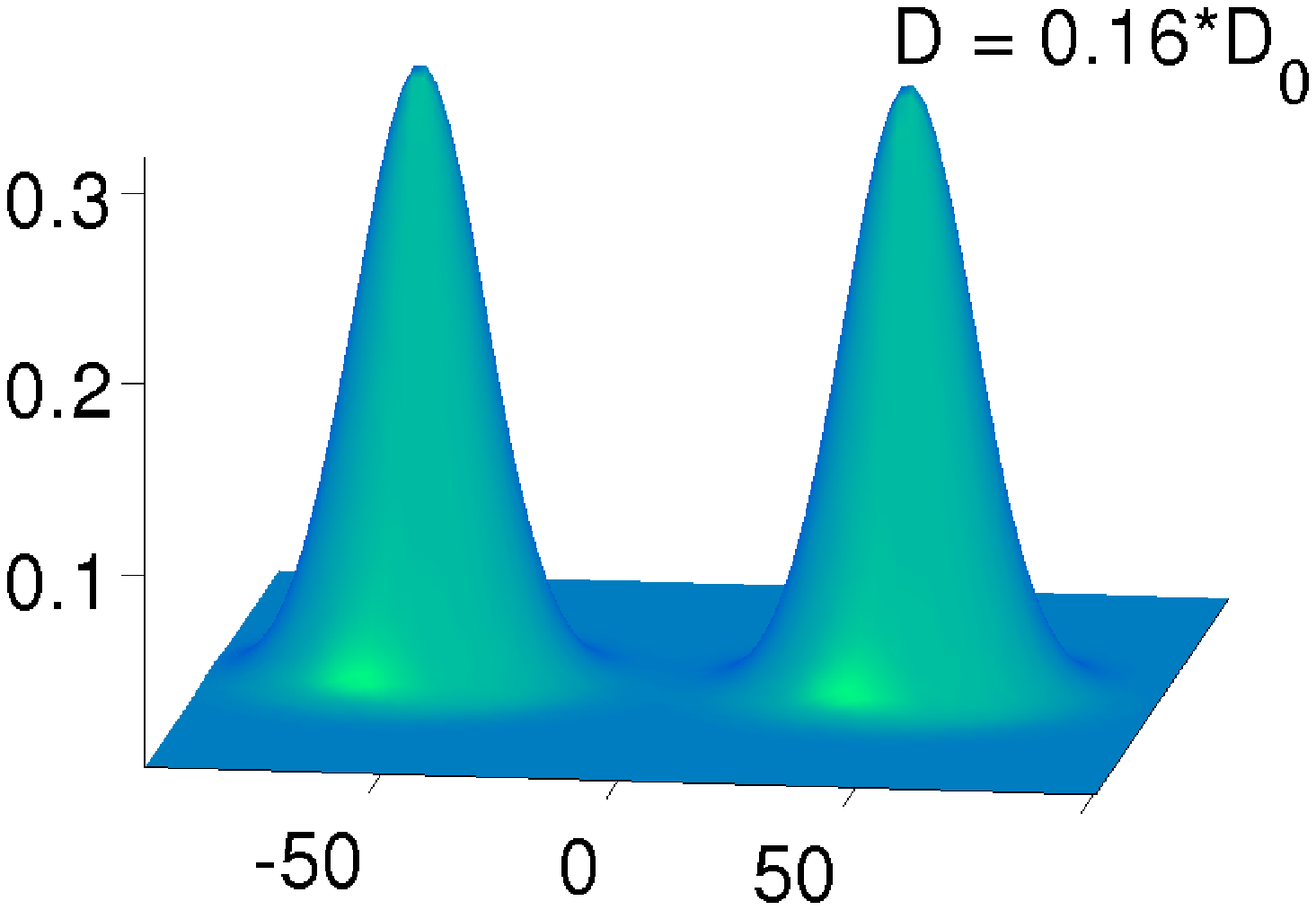}}
\subfigure{\includegraphics[width=.48\columnwidth]{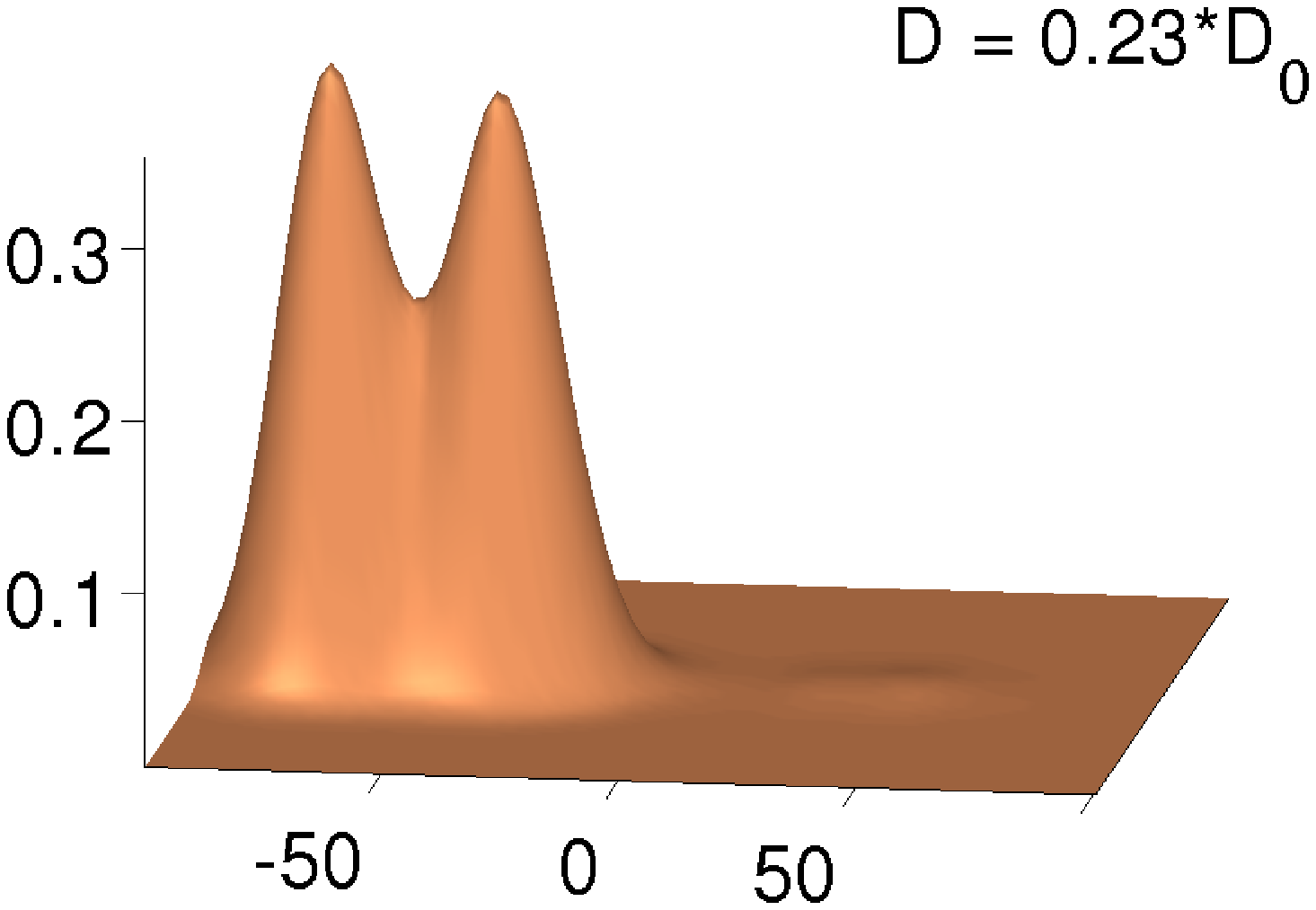}}
\caption{(Color online) The effect of the detuning strength $D$ on the behavior of the system. In the left: $D=0.16 D_0$, with $D_0=m^*\omega_0^2\times30$ nm. The upper plot shows the number of charges in left and right dots as a function of $a$ (see Fig. \ref{fig:electrons}).
The lower plot shows the singlet charge density at $a=100$ nm. In the right: $D=0.23 D_0$. The upper plot shows the number of charges a function of $a$ (see Fig. \ref{fig:electrons}). The lower plot shows the triplet
charge density at $a=100$ nm. The results are obtained adiabatically.
}
\label{fig:fail}
\end{figure}

The behavior of the singlet- and triplet-states was found to be quite sensitive to the strength of the detuning. If the detuning was too low, the singlet state would also split, and if it were too high, both the singlet and the
triplet would transfer to the left. The region in which the triplet would split and the singlet would not was found to be quite narrow, from $D=0.17 D_0$ to $D=0.22D_0$, with $D_0=m^*\omega_0^2\times30$nm ($a$ was increased from 0 nm to 100 nm).
Cases of too low and too high detunings are shown in Fig. \ref{fig:fail}. The results are obtained adiabatically. In the left plots, the detuning is $D=0.16D_0$ and in the right plots it is $D=0.23D_0$. In the $D=0.16D_0$ case, the singlet
is split at $a=100$ nm, although it seems to be transferring initially to the left as in Figs. \ref{fig:electrons} and \ref{fig:density}. In the $D=0.23D_0$ case, the triplet also localizes to the left.

The effect of the detuning can be analyzed qualitatively by looking at the energy levels of the system. As $a$ increases, the dots slide up (the right dot) or down (the left dot) the linear slope. The linear detuning lowers the potential of the left dot by a value $V_D(a)$ that increases with $a$.
In the split state, when one electron inhabits each dot, and the distance of the dots is large, the electrons can be thought
as non-interacting. The electrons stay split until the detuning lowers the potential energy of the left dot enough to overcome the Coulomb repulsion between them, i.e. the distance of the dots become high enough.
As the triplet is higher in energy than the singlet, there exists values of $a$ and $D$ so that the triplet is split while the singlet is not.

However, when $a$ is increased enough, the slope will eventually win in both the singlet- and the triplet-cases.
This might not hold anymore for very large displacements $a$ though, as the middle ridge may become too high for the electrons to tunnel through it. In real experiments,
the dot distance (the maximum distance of the Y-branches) is naturally limited to some maximum value, and the detuning can be set accordingly.

The dimensions of the dots, the values of $\alpha$ and $\beta$ in Eq. (\ref{eq:vext}), also affect the behavior of the system. If $\beta<\alpha$, it is preferable for the electrons
to align along the $y$-axis (see Fig. \ref{fig:yjunction}) in the initial dot at $t<t_1$, which results in both the singlet and the triplet transferring to the left.
Consequently, only values $\alpha<\beta$ were studied. 
These values affect the energy levels of the system and thus also the behavior of the electrons in the Y-junction. If $\alpha<\beta$, the detuning could be set so that only the triplet would be split at the end of the simulation with $a=100$ nm. 

In summary, we predict that a Y-shaped junction with an electrostatic detuning potential can be used to differentiate the two-body singlet- and triplet states in SAW induced electron transport. If the detuning strength
is suitable, the triplet states are split into the two Y-branches due to their repulsive exchange force, while the singlet state is preserved and is transferred into the energetically preferable branch.
This phenomenon could be used in quantum information technology. For example in quantum computing, it could allow measurement and control of two-electron qubits.

\begin{acknowledgments}
This research has been supported by the Academy of Finland through its Centres of Excellence Program (project no. 251748).
\end{acknowledgments}
\bibliography{perse}

\begin{thebibliography}{10}%
\makeatletter
\providecommand \@ifxundefined [1]{%
 \ifx #1\undefined \expandafter \@firstoftwo
 \else \expandafter \@secondoftwo
\fi
}%
\providecommand \@ifnum [1]{%
 \ifnum #1\expandafter \@firstoftwo
 \else \expandafter \@secondoftwo
\fi
}%
\providecommand \enquote [1]{``#1''}%
\providecommand \bibnamefont  [1]{#1}%
\providecommand \bibfnamefont [1]{#1}%
\providecommand \citenamefont [1]{#1}%
\providecommand\href[0]{\@sanitize\@href}%
\providecommand\@href[1]{\endgroup\@@startlink{#1}\endgroup\@@href}%
\providecommand\@@href[1]{#1\@@endlink}%
\providecommand \@sanitize [0]{\begingroup\catcode`\&12\catcode`\#12\relax}%
\@ifxundefined \pdfoutput {\@firstoftwo}{%
 \@ifnum{\z@=\pdfoutput}{\@firstoftwo}{\@secondoftwo}%
}{%
 \providecommand\@@startlink[1]{\leavevmode\special{html:<a href="#1">}}%
 \providecommand\@@endlink[0]{\special{html:</a>}}%
}{%
 \providecommand\@@startlink[1]{%
  \leavevmode
  \pdfstartlink
   attr{/Border[0 0 1 ]/H/I/C[0 1 1]}%
   user{/Subtype/Link/A<</Type/Action/S/URI/URI(#1)>>}%
  \relax
 }%
 \providecommand\@@endlink[0]{\pdfendlink}%
}%
\providecommand \url  [0]{\begingroup\@sanitize \@url }%
\providecommand \@url [1]{\endgroup\@href {#1}{\urlprefix}}%
\providecommand \urlprefix [0]{URL }%
\providecommand \Eprint[0]{\href }%
\@ifxundefined \urlstyle {%
  \providecommand \doi [1]{doi:\discretionary{}{}{}#1}%
}{%
  \providecommand \doi [0]{doi:\discretionary{}{}{}\begingroup
  \urlstyle{rm}\Url }%
}%
\providecommand \doibase [0]{http://dx.doi.org/}%
\providecommand \Doi[1]{\href{\doibase#1}}%
\providecommand \bibAnnote [3]{%
  \BibitemShut{#1}%
  \begin{quotation}\noindent
    \textsc{Key:}\ #2\\\textsc{Annotation:}\ #3%
  \end{quotation}%
}%
\providecommand \bibAnnoteFile [2]{%
  \IfFileExists{#2}{\bibAnnote {#1} {#2} {\input{#2}}}{}%
}%
\providecommand \typeout [0]{\immediate \write \m@ne }%
\providecommand \selectlanguage [0]{\@gobble}%
\providecommand \bibinfo [0]{\@secondoftwo}%
\providecommand \bibfield [0]{\@secondoftwo}%
\providecommand \translation [1]{[#1]}%
\providecommand \BibitemOpen[0]{}%
\providecommand \bibitemStop [0]{}%
\providecommand \bibitemNoStop [0]{.\EOS\space}%
\providecommand \EOS [0]{\spacefactor3000\relax}%
\providecommand \BibitemShut [1]{\csname bibitem#1\endcsname}%
\bibitem{loss}%
  \BibitemOpen
  \bibfield{author}{%
  \bibinfo {author} {\bibfnamefont{D.}~\bibnamefont{Loss}}\ and\ \bibinfo
  {author} {\bibfnamefont{D.~P.}\ \bibnamefont{DiVincenzo}},\ }%
  \bibfield{journal}{%
  \bibinfo {journal} {Phys. Rev. A}\ }%
  \textbf{\bibinfo {volume} {57}},\ \bibinfo {pages} {120} (\bibinfo {year}
  {1998})%
  \bibAnnoteFile{NoStop}{loss}%
\bibitem{levy}%
  \BibitemOpen
  \bibfield{author}{%
  \bibinfo {author} {\bibfnamefont{J.}~\bibnamefont{Levy}},\ }%
  \bibfield{journal}{%
  \bibinfo {journal} {Phys. Rev. Lett.}\ }%
  \textbf{\bibinfo {volume} {89}},\ \bibinfo {pages} {147902} (\bibinfo {year}
  {2002})%
  \bibAnnoteFile{NoStop}{levy}%
\bibitem{spins}%
  \BibitemOpen
  \bibfield{author}{%
  \bibinfo {author} {\bibfnamefont{R.}~\bibnamefont{Hanson}}, \bibinfo {author}
  {\bibfnamefont{L.~P.}\ \bibnamefont{Kouwenhoven}}, \bibinfo {author}
  {\bibfnamefont{J.~R.}\ \bibnamefont{Petta}}, \bibinfo {author}
  {\bibfnamefont{S.}~\bibnamefont{Tarucha}},\ and\ \bibinfo {author}
  {\bibfnamefont{L.~M.~K.}\ \bibnamefont{Vandersypen}},\ }%
  \bibfield{journal}{%
  \bibinfo {journal} {Rev. Mod. Phys.}\ }%
  \textbf{\bibinfo {volume} {79}},\ \bibinfo {pages} {1217} (\bibinfo {year}
  {2007})%
  \bibAnnoteFile{NoStop}{spins}%
\bibitem{petta}%
  \BibitemOpen
  \bibfield{author}{%
  \bibinfo {author} {\bibfnamefont{J.~R.}\ \bibnamefont{Petta}}, \bibinfo
  {author} {\bibfnamefont{A.~C.}\ \bibnamefont{Johnson}}, \bibinfo {author}
  {\bibfnamefont{J.~M.}\ \bibnamefont{Taylor}}, \bibinfo {author}
  {\bibfnamefont{E.~A.}\ \bibnamefont{Laird}}, \bibinfo {author}
  {\bibfnamefont{A.}~\bibnamefont{Yacoby}}, \bibinfo {author}
  {\bibfnamefont{M.~D.}\ \bibnamefont{Lukin}}, \bibinfo {author}
  {\bibfnamefont{C.~M.}\ \bibnamefont{Marcus}}, \bibinfo {author}
  {\bibfnamefont{M.~P.}\ \bibnamefont{Hanson}},\ and\ \bibinfo {author}
  {\bibfnamefont{A.~C.}\ \bibnamefont{Gossard}},\ }%
  \bibfield{journal}{%
  \bibinfo {journal} {Science}\ }%
  \textbf{\bibinfo {volume} {309}},\ \bibinfo {pages} {2180} (\bibinfo {year}
  {2005})%
  \bibAnnoteFile{NoStop}{petta}%
\bibitem{sarkka}%
  \BibitemOpen
  \bibfield{author}{%
  \bibinfo {author} {\bibfnamefont{J.}~\bibnamefont{S{\"a}rkk{\"a}}}\ and\
  \bibinfo {author} {\bibfnamefont{A.}~\bibnamefont{Harju}},\ }%
  \bibfield{journal}{%
  \bibinfo {journal} {New J. Phys.}\ }%
  \textbf{\bibinfo {volume} {13}},\ \bibinfo {pages} {043010} (\bibinfo {year}
  {2011})%
  \bibAnnoteFile{NoStop}{sarkka}%
\bibitem{petta2}%
  \BibitemOpen
  \bibfield{author}{%
  \bibinfo {author} {\bibfnamefont{J.~R.}\ \bibnamefont{Petta}}, \bibinfo
  {author} {\bibfnamefont{H.}~\bibnamefont{Lu}},\ and\ \bibinfo {author}
  {\bibfnamefont{A.~C.}\ \bibnamefont{Gossard}},\ }%
  \bibfield{journal}{%
  \bibinfo {journal} {Science}\ }%
  \textbf{\bibinfo {volume} {327}},\ \bibinfo {pages} {669} (\bibinfo {year}
  {2010})%
  \bibAnnoteFile{NoStop}{petta2}%
\bibitem{studenkin}%
  \BibitemOpen
  \bibfield{author}{%
  \bibinfo {author} {\bibfnamefont{S.~A.}\ \bibnamefont{Studenikin}}, \bibinfo
  {author} {\bibfnamefont{G.~C.}\ \bibnamefont{Aers}}, \bibinfo {author}
  {\bibfnamefont{G.}~\bibnamefont{Granger}}, \bibinfo {author}
  {\bibfnamefont{L.}~\bibnamefont{Gaudreau}}, \bibinfo {author}
  {\bibfnamefont{A.}~\bibnamefont{Kam}}, \bibinfo {author}
  {\bibfnamefont{P.}~\bibnamefont{Zawadzki}}, \bibinfo {author}
  {\bibfnamefont{Z.~R.}\ \bibnamefont{Wasilewski}},\ and\ \bibinfo {author}
  {\bibfnamefont{A.~S.}\ \bibnamefont{Sachrajda}},\ }%
  \bibfield{journal}{%
  \bibinfo {journal} {Phys. Rev. Lett.}\ }%
  \textbf{\bibinfo {volume} {108}},\ \bibinfo {pages} {226802} (\bibinfo {year}
  {2012})%
  \bibAnnoteFile{NoStop}{studenkin}%
\bibitem{bluhm}%
  \BibitemOpen
  \bibfield{author}{%
  \bibinfo {author} {\bibfnamefont{H.}~\bibnamefont{Bluhm}}, \bibinfo {author}
  {\bibfnamefont{S.}~\bibnamefont{Foletti}}, \bibinfo {author}
  {\bibfnamefont{I.}~\bibnamefont{Neder}}, \bibinfo {author}
  {\bibfnamefont{M.}~\bibnamefont{Rudner}}, \bibinfo {author}
  {\bibfnamefont{D.}~\bibnamefont{Mahalu}}, \bibinfo {author}
  {\bibfnamefont{V.}~\bibnamefont{Umansky}},\ and\ \bibinfo {author}
  {\bibfnamefont{A.}~\bibnamefont{Yacoby}},\ }%
  \bibfield{journal}{%
  \bibinfo {journal} {Nature Physics}\ }%
  \textbf{\bibinfo {volume} {7}},\ \bibinfo {pages} {109} (\bibinfo {year}
  {2011})%
  \bibAnnoteFile{NoStop}{bluhm}%
\bibitem{shulman}%
  \BibitemOpen
  \bibfield{author}{%
  \bibinfo {author} {\bibfnamefont{M.~D.}\ \bibnamefont{Shulman}}, \bibinfo
  {author} {\bibfnamefont{O.~E.}\ \bibnamefont{Dial}}, \bibinfo {author}
  {\bibfnamefont{S.~P.}\ \bibnamefont{Harvey}}, \bibinfo {author}
  {\bibfnamefont{H.}~\bibnamefont{Bluhm}}, \bibinfo {author}
  {\bibfnamefont{V.}~\bibnamefont{Umansky}},\ and\ \bibinfo {author}
  {\bibfnamefont{A.}~\bibnamefont{Yacoby}},\ }%
  \bibfield{journal}{%
  \bibinfo {journal} {Science}\ }%
  \textbf{\bibinfo {volume} {336}},\ \bibinfo {pages} {202} (\bibinfo {year}
  {2012})%
  \bibAnnoteFile{NoStop}{shulman}%
\bibitem{barnes}%
  \BibitemOpen
  \bibfield{author}{%
  \bibinfo {author} {\bibfnamefont{C.~H.~W.}\ \bibnamefont{Barnes}}, \bibinfo
  {author} {\bibfnamefont{J.~M.}\ \bibnamefont{Shilton}},\ and\ \bibinfo
  {author} {\bibfnamefont{A.~M.}\ \bibnamefont{Robinson}},\ }%
  \bibfield{journal}{%
  \bibinfo {journal} {Phys. Rev. B}\ }%
  \textbf{\bibinfo {volume} {62}},\ \bibinfo {pages} {8410} (\bibinfo {year}
  {2000})%
  \bibAnnoteFile{NoStop}{barnes}%
\bibitem{saw3}%
  \BibitemOpen
  \bibfield{author}{%
  \bibinfo {author} {\bibfnamefont{R.~P.~G.}\ \bibnamefont{McNeil}}, \bibinfo
  {author} {\bibfnamefont{M.}~\bibnamefont{Kataoka}}, \bibinfo {author}
  {\bibfnamefont{C.~J.~B.}\ \bibnamefont{Ford}}, \bibinfo {author}
  {\bibfnamefont{C.~H.~W.}\ \bibnamefont{Barnes}}, \bibinfo {author}
  {\bibfnamefont{D.}~\bibnamefont{Anderson}}, \bibinfo {author}
  {\bibfnamefont{G.~A.~C.}\ \bibnamefont{Jones}}, \bibinfo {author}
  {\bibfnamefont{I.}~\bibnamefont{Farrer}},\ and\ \bibinfo {author}
  {\bibfnamefont{D.~A.}\ \bibnamefont{Ritchie}},\ }%
  \bibfield{journal}{%
  \bibinfo {journal} {Nature}\ }%
  \textbf{\bibinfo {volume} {477}},\ \bibinfo {pages} {439} (\bibinfo {year}
  {2011})%
  \bibAnnoteFile{NoStop}{saw3}%
\bibitem{saw4}%
  \BibitemOpen
  \bibfield{author}{%
  \bibinfo {author} {\bibfnamefont{S.}~\bibnamefont{Hermelin}}, \bibinfo
  {author} {\bibfnamefont{S.}~\bibnamefont{Takada}}, \bibinfo {author}
  {\bibfnamefont{M.}~\bibnamefont{Yamamoto}}, \bibinfo {author}
  {\bibfnamefont{S.}~\bibnamefont{Tarucha}}, \bibinfo {author}
  {\bibfnamefont{A.~D.}\ \bibnamefont{Wieck}}, \bibinfo {author}
  {\bibfnamefont{L.}~\bibnamefont{Saminadayar}}, \bibinfo {author}
  {\bibfnamefont{C.}~\bibnamefont{Baeuerle}},\ and\ \bibinfo {author}
  {\bibfnamefont{T.}~\bibnamefont{Meunier}},\ }%
  \bibfield{journal}{%
  \bibinfo {journal} {Nature}\ }%
  \textbf{\bibinfo {volume} {477}},\ \bibinfo {pages} {435} (\bibinfo {year}
  {2011})%
  \bibAnnoteFile{NoStop}{saw4}%
\bibitem{sawy}%
  \BibitemOpen
  \bibfield{author}{%
  \bibinfo {author} {\bibfnamefont{V.}~\bibnamefont{Talyanskii}}, \bibinfo
  {author} {\bibfnamefont{M.}~\bibnamefont{Graham}},\ and\ \bibinfo {author}
  {\bibfnamefont{H.}~\bibnamefont{Beere}},\ }%
  \bibfield{journal}{%
  \bibinfo {journal} {Appl. Phys. Lett.}\ }%
  \textbf{\bibinfo {volume} {88}} (\bibinfo {year} {2006})%
  \bibAnnoteFile{NoStop}{sawy}%
\bibitem{kataoka}%
  \BibitemOpen
  \bibfield{author}{%
  \bibinfo {author} {\bibfnamefont{M.}~\bibnamefont{Kataoka}}, \bibinfo
  {author} {\bibfnamefont{M.~R.}\ \bibnamefont{Astley}}, \bibinfo {author}
  {\bibfnamefont{A.~L.}\ \bibnamefont{Thorn}}, \bibinfo {author}
  {\bibfnamefont{D.~K.~L.}\ \bibnamefont{Oi}}, \bibinfo {author}
  {\bibfnamefont{C.~H.~W.}\ \bibnamefont{Barnes}}, \bibinfo {author}
  {\bibfnamefont{C.~J.~B.}\ \bibnamefont{Ford}}, \bibinfo {author}
  {\bibfnamefont{D.}~\bibnamefont{Anderson}}, \bibinfo {author}
  {\bibfnamefont{G.~A.~C.}\ \bibnamefont{Jones}}, \bibinfo {author}
  {\bibfnamefont{I.}~\bibnamefont{Farrer}}, \bibinfo {author}
  {\bibfnamefont{D.~A.}\ \bibnamefont{Ritchie}},\ and\ \bibinfo {author}
  {\bibfnamefont{M.}~\bibnamefont{Pepper}},\ }%
  \bibfield{journal}{%
  \bibinfo {journal} {Phys. Rev. Lett.}\ }%
  \textbf{\bibinfo {volume} {102}},\ \bibinfo {pages} {156801} (\bibinfo {year}
  {2009})%
  \bibAnnoteFile{NoStop}{kataoka}%
\end{thebibliography}%

\end{document}